\documentclass[twocolumn,twocolappendix]{aastex631}

\usepackage{amsmath}
\usepackage{relsize}

\DeclareMathOperator{\arctantwo}{arctan2}

\newcommand{\change}[1]{{#1}}

\newcommand{\CfA}{\affiliation{Center for Astrophysics | Harvard \& Smithsonian, 60 Garden St., Cambridge, MA 02138, USA}}

\usepackage{url}
\usepackage{footnote}

\begin{document}

\title{Detection of knots in the supernova type Iax remnant Pa\,30}

\author[0000-0000-0000-0001]{Tim Cunningham$^{\dagger}$}
\altaffiliation{These authors contributed equally.}
\affiliation{Center for Astrophysics | Harvard \& Smithsonian, 60 Garden St., Cambridge, MA 02138, USA}
\affiliation{Department of Physics, University of Warwick, Coventry CV4 7AL, UK}
\thanks{NASA Hubble Fellow}
\email{tim.cunningham@cfa.harvard.edu}

\author[0000-0002-4770-5388]{Ilaria Caiazzo}
\altaffiliation{These authors contributed equally.}
\affiliation{Institute of Science and Technology Austria, Am Campus 1, 3400 Klosterneuburg, Austria}

\author[0000-0002-7868-1622]{John C. Raymond}
\CfA

\author[0000-0003-0214-609X]{Scott J. Kenyon}
\CfA

\shorttitle{Detection of Knots in Pa 30}
\shortauthors{Cunningham-Caiazzo et al.}

\begin{abstract}
We present deep, narrowband Gemini North/GMOS images of the Pa\,30 nebula that reveal rich small‑scale structure in the ejecta filaments. In [S\,{\sc ii}], the number of detected filaments increases by about an order of magnitude compared to previous studies. The imaging further reveals that the filaments \change{are not smooth radial streaks, but instead are composed of cascades of knots or clumps}. The images in [O\,{\sc iii}] yield the first confirmation that previously detected diffuse [O\,{\sc iii}] emission in fact originates from the same [S\,{\sc ii}] filaments.
We measure a nebula center consistent with the position of the central star, indicating that the remnant has not received a large transverse kick. The 3$\sigma$ upper limit on the velocity of the stellar kick is $\lesssim$\,100\,km/s .
Cross‑sectional profiles of the filaments in [S\,{\sc ii}] and [O\,{\sc iii}] show a central peak on top of a nebular background, the latter being likely due to diffuse emission or unresolved filaments. The median [S\,{\sc ii}] width of $0.7''$ is slightly broader than the $0.50''$ seeing, providing some evidence of partially resolved filaments \change{and implying a characteristic knot diameter of ${\sim}10^{16}$\,cm}. The $\approx$1'' widths measured in [O\,{\sc iii}] are consistent with the poorer seeing of the oxygen images. \change{The discovery of this clumpy substructure provides a new observational constraint on models for the origin and evolution of the remarkable filamentary ejecta of Pa\,30.}
\end{abstract}

\keywords{Type Ia supernovae (1728);  supernova remnants (1667)}

\section{Introduction} \label{sec:intro}
The supernova (SN) observed from Earth in 1181 was recorded by Japanese, Chinese, and Arabic astronomers \citep{schaefer2023,fischer2025}, but it was only in 2013 that a citizen science campaign led to the discovery of its associated remnant \citep{ritter2021}. The initial identification entailed the discovery of a previously unidentified nebula in archival imaging in the W3 (12\,$\mu$m) and W4 (20\,$\mu$m) filters of the \textit{Wide-field Infrared Survey Explorer} \citep[\textit{WISE;}][]{wise2010}. Follow-up narrowband imaging in [O\,{\sc iii}] obtained by \citet{ritter2021} on the 2.1 m Kitt Peak National Observatory (KPNO) telescope revealed faint, diffuse nebular emission. Further follow-up narrowband imaging in [S\,{\sc ii}] from the 2.4\,m telescope at the MDM Observatory at Kitt Peak, Arizona was presented by \citet{fesen2023}. These observations revealed remarkable radial filaments threading the nebula with a high level of circular symmetry. Optical spectroscopy of the nebula reveals a velocity distribution of the [S\,{\sc ii}] emission peaking at 1000\,km/s \citep{ritter2021,cunningham2024}, with maximum velocities reaching $\approx$1400 km/s \citep{cunningham2024}. 
At the center of the remnant is a hot star (IRAS\,00500+6713) with an effective temperature of $\approx$200,000\,K and broad high-ionization emission lines that provide evidence of a fast stellar wind of $\approx$16,000\,km/s \citep{gvaramadze2019,lykou2023}.

Pa\,30 is unique among all known supernova remnants.
The low peak luminosity inferred from historical records \citep{schaefer2023}, together with the modest ejecta velocities observed in the remnant and the presence of a surviving central star, has led to the suggestion that SN\,1181 belongs to the class of subluminous Type Ia supernovae known as Type Iax \citep{foley2013, oskinova2020, ritter2021, lykou2023, schaefer2023, fesen2023}. This interpretation is further supported by the absence of hydrogen and helium in both the stellar atmosphere and the surrounding nebula, as well as by the detection of neon, magnesium, silicon, and sulfur in the X-ray spectra of the central star and nebula \citep{oskinova2020}, all of which are consistent with a relatively low-energy thermonuclear explosion \citep{ko2024,shao2025}. 

There is growing evidence that Type Iax supernovae may arise from multiple progenitor channels, with some events originating from double white dwarf mergers \citep{karambelkar2021} and others from white dwarfs in binary systems with non-degenerate stellar companions \citep{mccully2014,foley2014}. Simulations have predicted that the remnant star might receive a kick, with velocities in the range 6--600\,km/s \citep{jordan2012,kromer2013,fink2014,lach2022}, while the discovery of high-velocity ($>500$\,km\,s$^{-1}$) candidate remnants suggests stronger kicks might be possible \citep{vennes2017,raddi2017,raddi2018a,raddi2018b,raddi2019,el-badry2023,bhat2026}. In the case of Pa\,30, it has been proposed that the supernova originated from the merger of two white dwarfs \citep{gvaramadze2019,yao2023,ko2026,piro2026,uno2026}, while the powerful outflows observed from the surviving central star are thought to be accelerated by magnetocentrifugal flinging \citep{zhong2024,ko2024}. The symmetry of the nebula and the low proper motion of the central star imply a low kick velocity, but no quantitative constraints have yet been made.

The morphology of the Pa\,30 nebula differs markedly from the clumpy structures typically seen in Type Ia remnants, which are usually shaped by Rayleigh–Taylor and Kelvin–Helmholtz instabilities. One proposed explanation for the filaments is that the fast wind from the central star accelerates low-density ejecta into filamentary structures through Rayleigh–Taylor instabilities \citep{fesen2023,ko2024}, as seen in the wind-blown tails of novae like GK Persei and DQ Herculis \citep{vaytet2007,shara2012}, and of some planetary nebulae \citep{odell1996,wesson2024}. Alternative models suggest that the filaments form through instabilities at the interface between the supernova ejecta and circumstellar material (CSM), with either efficient cooling or high density contrast suppressing turbulence and producing the observed nearly ballistic radial structures \citep{duffell2024,pikus2025,coughlin2025}.

We here report deep, narrowband imaging with [S\,{\sc ii}] and [O\,{\sc iii}] that reveals knots, or clumps, in the [S\,{\sc ii}] nebula.
In Section\,\ref{sec:obs}, we describe the observations and data reduction. In Section\,\ref{sec:results} we present an analysis of the images, including a characterisation of the small-scale and global structure of the nebula, and present general properties of the filaments. In Section\,\ref{sec:summary}, we present a summary of the findings.

\section{Observations} \label{sec:obs}
We obtained 6.9 hours of Gemini Multi-Object Spectrograph \citep[GMOS;][]{hook2004-gmos} imaging from the Gemini North telescope through the Gemini Fast Turnaround program (ID GN-2025A-FT-112; PI: Cunningham). The observations comprised deep imaging of the Pa\,30 nebula, making use of two narrowband filters: [S\,{\sc ii}] and [O\,{\sc iii}]. We obtained 2 hours of imaging in the [S\,{\sc ii}] filter G0317 with central wavelength 6720\,\AA\ and effective transmission from 6694--6737\,\AA. The [S\,{\sc ii}] imaging was awarded observations with optimal image quality (IQ20). The [O\,{\sc iii}] imaging was obtained with the G0318 filter, with central wavelength 4990\,\AA\ and effective transimission from 4965--5015\,\AA. These observations were awarded with 1 hour of IQ20, and the remaining 3 hours with IQ70. An analysis of $\approx$120 stars in the field finds an empirical point-spread function of FWHM\,$=$\,0.50\,$\pm$\,0.01 and 0.85\,$\pm$\,0.02'', for the final mosaicked [S\,{\sc ii}] and [O\,{\sc iii}] images, respectively. Observations were carried out in queue mode across six nights between July 26, 2025 and August 05, 2025. The observations were dithered to close the gap between CCDs, essential because the on-sky width of the central CCD is comparable to the diameter of the nebula. Data were reduced using the standard GMOS reduction pipeline DRAGONS \citep{labrie2023-dragons} Version 4.2.1. The processing included overscan subtraction, trimming, bias subtraction, flat-field correction, gain normalization, and mosaicking of the three CCDs, followed by astrometric calibration, sky background subtraction, and stacking of the individual exposures to produce the final science frames.  This process was carried out for both the [S\,{\sc ii}] and [O\,{\sc iii}] observations. Finally, the [O\,{\sc iii}] image was registered against the [S\,{\sc ii}] image to make a final fine astrometric correction.

\begin{figure*}
	\centering    
\includegraphics[width=1.0\textwidth]{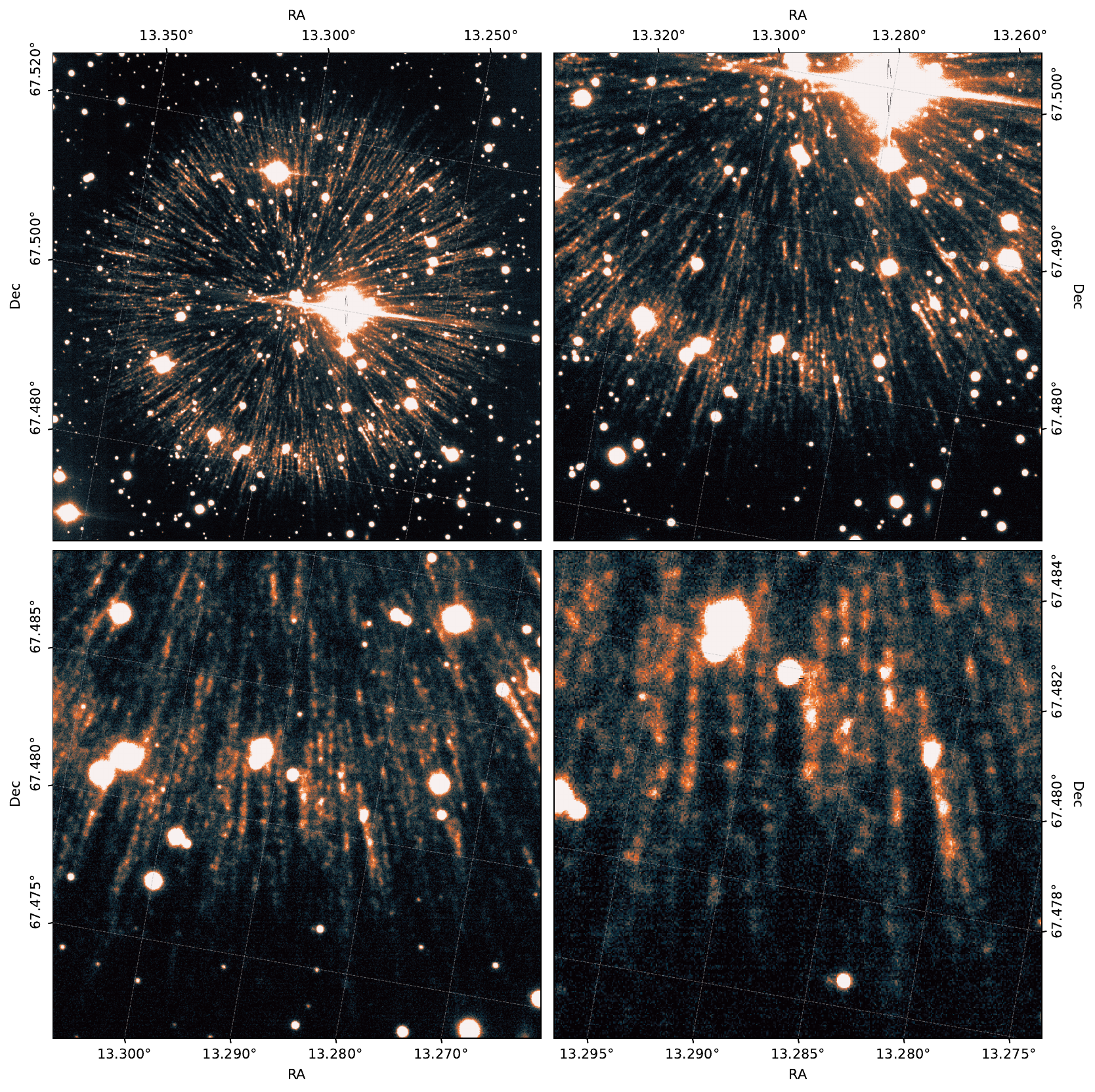} 
\caption{Narrowband imaging of the Pa\,30 nebula obtained with Gemini GMOS with the [S\,{\sc ii}] filter. The imaging reveals a high filling factor of filaments. In general the filaments are a cascade of knots. The grid spacing in the lower-right panel is 18'', illustrating a characteristic knot spacing of $\approx$1–6'' along the radial extent of the filaments.}
	\label{fg:ql_4panel_south}
\end{figure*}

\begin{figure*}
	\centering    
\includegraphics[width=1.0\textwidth]{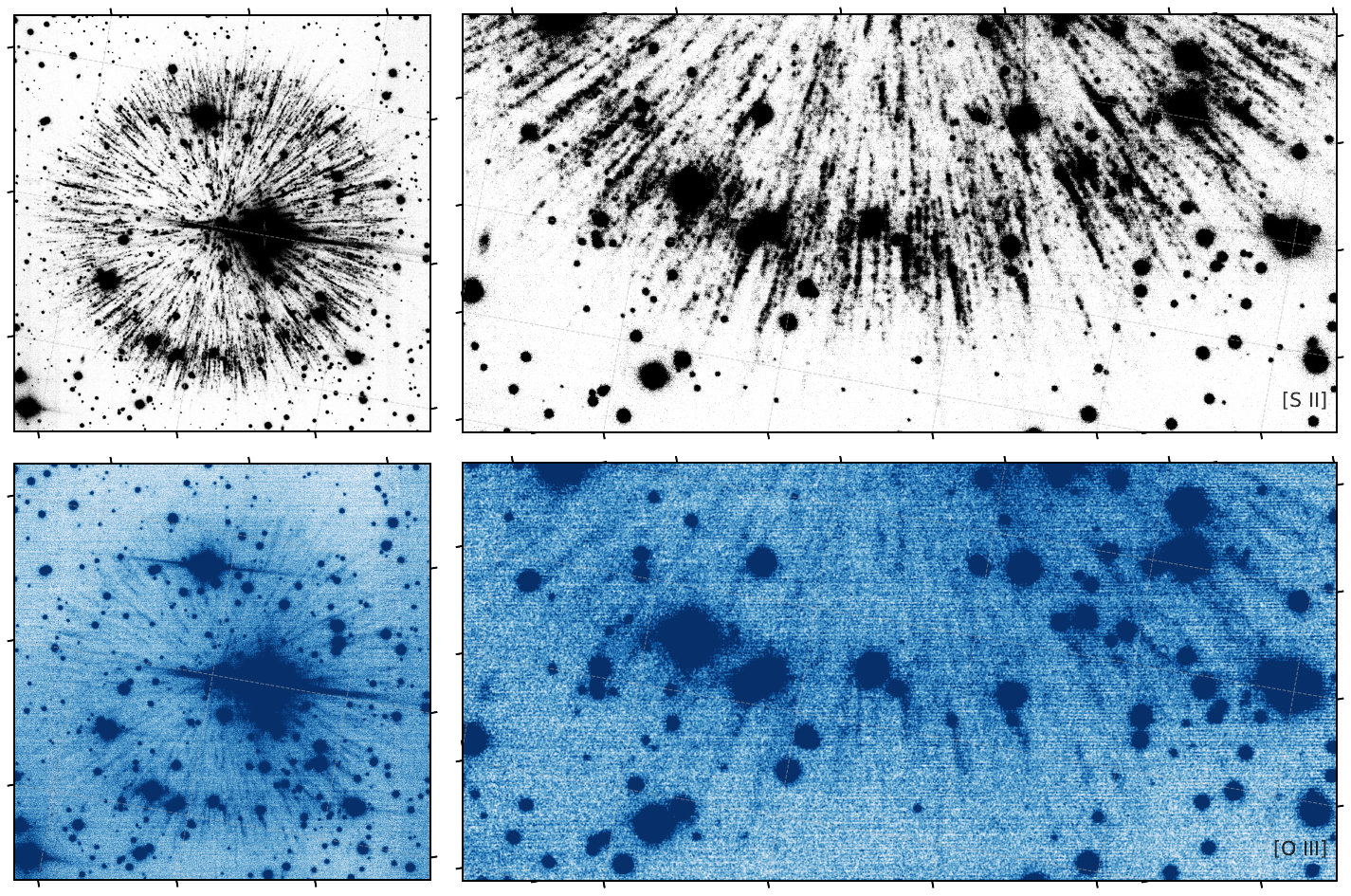} 
\caption{Narrowband imaging of the Pa\,30 nebula obtained with Gemini GMOS with the [S\,{\sc ii}] filter (upper panels) and [O\,{\sc iii}] filter (lower panels). The upper and lower panels are astrometrically aligned (i.e., the same field of view). The [O\,{\sc iii}] filaments are cospatial with the [S\,{\sc ii}] filaments.}
	\label{fg:ql_oiii_4panel}
\end{figure*}

\section{Results}
\label{sec:results}
\subsection{Filamentary strings of knots}
\label{sec:knots}
Figure~\ref{fg:ql_4panel_south} presents the [S\,{\sc ii}] image (upper-left panel) along with several zoomed-in views of the southern region of the nebula. Figure~\ref{fg:ql_oiii_4panel} compares the [S\,{\sc ii}] image (upper panels) with the [O\,{\sc iii}] image (lower panels). A visual inspection of the new images reveals that (1) the number of filaments is at least an order of magnitude greater than previously detected in earlier images \citep[e.g.][]{fesen2023}; (2) the radial filaments are not smooth, but instead consist of chains of tiny knots near the resolution limit in the [S\,{\sc ii}] image ($0.5$''); and (3) the [O\,{\sc iii}] emission exhibits the same filamentary structure as the [S\,{\sc ii}] emission. \change{In Appendix Section~\ref{sec:cross-correlation}, we show that filaments in the [O\,{\sc iii}] image are co-spatial with those in the [S\,{\sc ii}] image. The signal-to-noise in the [O\,{\sc iii}] image, however, is too low to explore whether the ratio between [S\,{\sc ii}] and [O\,{\sc iii}] is constant or varies along or across the filaments. Such analysis would require either deeper [O\,{\sc iii}] imaging, or a high angular resolution integrated field unit spectroscopy campaign.}

The separation and frequency of the knots are remarkably similar along each filament and across different filaments. 
As we show in section~\ref{sec:profile}, the knots are marginally resolved, with an average \change{angular size of ${\approx}\,0.6''$ (or $\approx$0.7\% of the radius of the nebula).} Adopting a distance of $D\,{=}\,2.3\,{\pm}\,0.1$\,kpc to the central star of the Pa\,30 nebula \citep{gaia2021}, we therefore take $10^{16}$ cm as their characteristic scale.  Somewhat similar strings of knots can be seen in the core collapse supernova remnant Cas A \citep{milisavljevic2024, rho2024}, although without the ordered radial distribution.  In both cases, the knots are likely to be formed by some combination of Kelvin-Helmholtz and thermal instabilities acting on narrow filaments subject to shear flows.  Three basic pictures for models of the filamentary structure have been suggested, and we consider how knots might arise in these scenarios.

\smallskip
\noindent
{\it Interaction with central star wind:}  \cite{fesen2023} proposed that the 16,000 km s$^{-1}$ wind from the central star \citep{gvaramadze2019} encounters clumps in a slower shell of ejecta from the explosion, and that the wind ablates the clumps to form radial filaments.  This picture does not naturally explain the ballistic motion of the filaments \citep{cunningham2024}, in which the filaments expand with velocities proportional to their distances from the central star, and their time of origin coincides with the observed SN event in 1181 \citep{ritter2021}. Also, the quasi-regular frequency of knots along the radial extent of the filaments is in contrast with the expected morphology for evaporating clumps.  We will not discuss this picture further.

\smallskip
\noindent
{\it Rayleigh-Taylor instability with strong radiative cooling:} \cite{duffell2024} and \cite{pikus2025} present models in which dense SN ejecta encounter the ambient interstellar medium (ISM).  The strong radiative cooling of the ejecta suppresses the Kelvin-Helmholtz instability at the tips of the Raleigh-Taylor fingers, resulting in very narrow RT fingers.  This model predicts very nearly ballistic expansion, provided that the interaction occurs early in the evolution.  Similar narrow fingers appeared in the simulation of the interaction of Cas A ejecta with its circumstellar medium by \cite{orlando2025}. 

In the models with high cooling rates shown by \cite{pikus2025}, the filaments are separated by unshocked ISM.  The densities in the filaments are 100-700 $\rm cm^{-3}$ \citep{fesen2023}, and \cite{pikus2025} assume an ambient density of about 0.1 $\rm cm^{-3}$, so the density contrast is of order 5000.  The growth rate of the Kelvin-Helmholtz instability can be estimated as

\begin{equation}
\gamma = k (\rho_1/\rho_2)^{1/2} (U_2-U_1)
\end{equation}

\noindent
where $k$ is the wavenumber, $U$ is the velocity, $\rho$ is the density, and the numbers 1 and 2 correspond to the ISM and the filaments, respectively. The large density contrast reduces the growth rate.  Taking $(\rho_1/\rho_2)$ = 1/5000 and a velocity difference of 1000 $\rm km~s^{-1}$, the growth time at scales around $10^{16}$ cm is about 160 years.  Thus, the \cite{duffell2024} and \cite{pikus2025} model is able to produce knots along the filaments by means of the Kelvin-Helmholtz instability, though the growth time is somewhat marginal.  

The cooling time of the filament is much shorter, around one month based on the cooling coefficients for silicate Earth composition taken from the rates for the abundant elements computed for SN ejecta \citep{raymond2018}. Therefore, thermal instability might accelerate the condensation of knots initiated by the Kelvin-Helmholtz instability.  It is also possible that thermal instability alone produces the knots, since radiatively cooling gas behind a shock is notoriously thermally unstable \citep{innes1992}, and the rapid cooling of sublimated dust will make the instability even stronger.  It is not clear whether thermal instability of a Rayleigh-Taylor plume would lead to a structure like the observed series of knots, or to chaotic small-scale structure.

\smallskip
\noindent
{\it \change{Rayleigh-Taylor instability from wind-CSM interaction:}}  \cite{coughlin2025} presented a simple 2D model of the Rayleigh Taylor instability between fluids with density contrasts of 2, 10 and 100, and they find that the high density contrast case yields very narrow Rayleigh-Taylor fingers, analogous to those seen in Pa 30.  They do not explain the high density contrast with radiative cooling, but instead posit an interaction between the circumstellar medium with a density of order 100 $\rm cm^{-3}$ and a wind launched by the merged white dwarf.  This wind would have higher density and lower speed than the wind observed today, and the interaction would occur at a distance of order $10^{15}~\rm cm$ from the star.  This model explains the lengths of the filaments in terms of the time when the density contrast and deceleration rate were large enough to drive the Rayleigh-Taylor instability.  Although the model does not have very much detail about the current state of the filaments, the gas between the filaments is taken to be shocked ambient material.  Therefore, the velocity contrast between the filaments and the gas between them is modest and the Kelvin-Helmholtz time scale is on the order of the age of the SNR.  Thus, it would be difficult for this model to account for the knots.  However, the knots could have formed by thermal instability, perhaps at an earlier stage of the evolution.  The \cite{coughlin2025} model does not include an energy equation, so it does not make a prediction about thermal instabilities.

\subsection{Rolling Hough Transform}
\label{sec:rht}
\begin{figure}
    \centering
    \includegraphics[width=0.8\columnwidth]{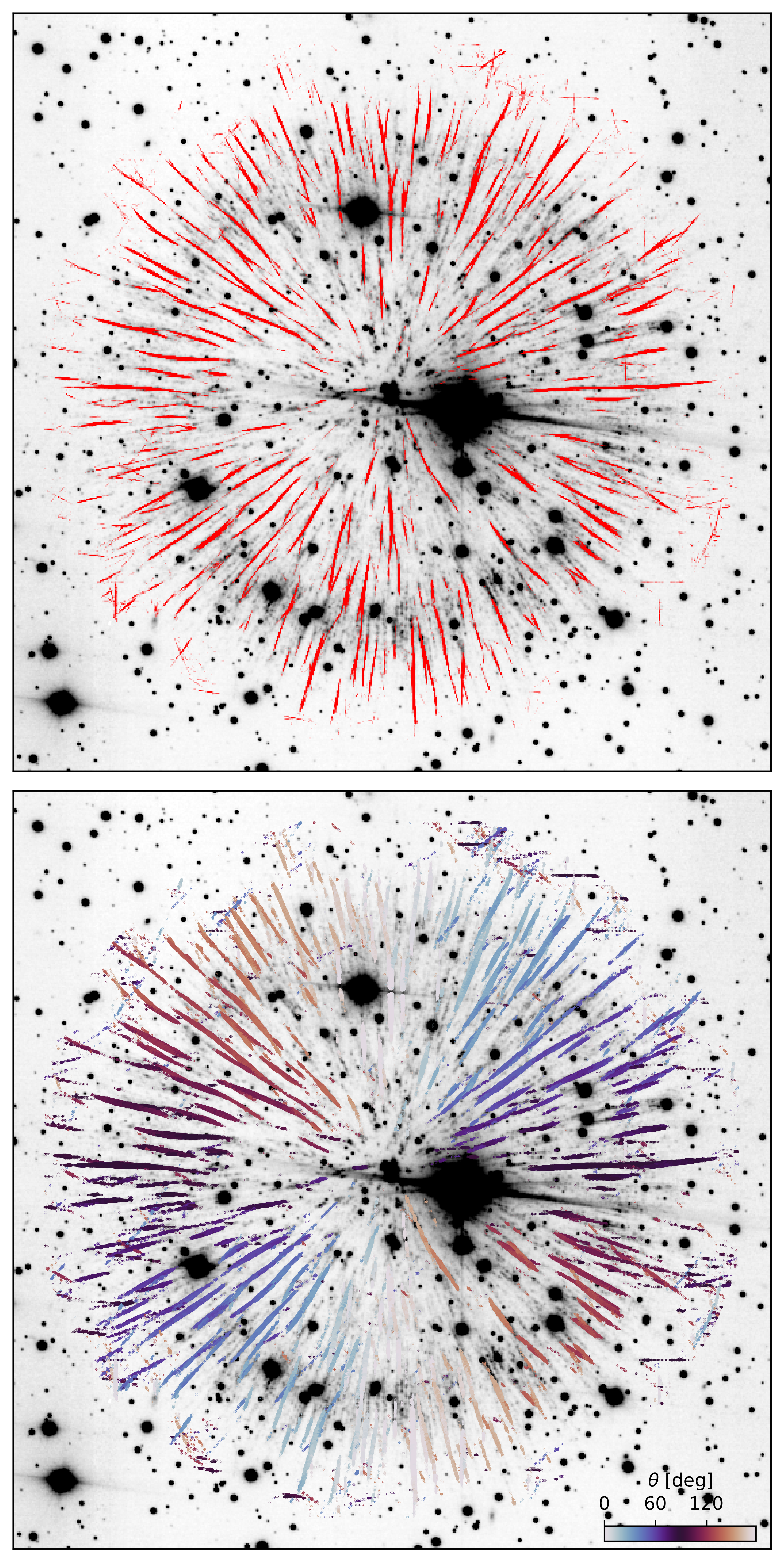}
    \caption{In greyscale, we show the [S\,{\sc ii}] image, smoothed to 0.4'', marginally below the seeing (0.5''). \textbf{Upper panel}: in red, we overlay the integrated RHT intensity for every pixel from an RHT analysis with parameters $D_{\rm w}\,{=}\,15''$, $r_{\rm s}\,{=}\,2''$, and $Z\,{=}\,0.7$. We plot the RHT intensities only up to 90'' from the central star, excluding four masking regions which were selected to avoid prominent diffraction spikes from two heavily saturated stars. \textbf{Lower panel}: The same RHT map as in the upper panel, but here the color of each pixel reflects the angle corresponding to the maximum intensity in the RHT spectrum. We can see that the RHT recovers the directionality of each filament.}
 	\label{fg:rht_overlay}
\end{figure}

\noindent
To quantify the structure and morphology of the remnant in the new images, we make use of the Rolling Hough Transform (RHT), a technique first developed for analysis of linear filaments in astrophysical imaging by \citet{clark2014-rht}. The RHT is a local, moving-window version of the classic Hough Transform \citep{hough1962} used to detect the orientation of linear structures in images. For the analysis, we use an open-source Python implementation of the algorithm \citep{clark2020-rht-code}. 

The parameters used in the implementations are the smoothing radius of the unsharp mask ($r_{\rm{s}}$), the window length ($D_{\rm{w}}$),  and a detection threshold ($Z$). The first determines the mask that is used in an initial high-pass Fourier filter on the image, and it should be close to 2--5 times the thickness of the linear structure. The RHT then computes, for each pixel in the image, the dominant orientation in a circular window of diameter $D_{\rm{w}}$ surrounding the pixel. Finally, the detection threshold is defined as the fraction of any given angle that must have some intensity to be counted. 

The top panel of Fig.\,\ref{fg:rht_overlay} shows an example of the RHT when applied to the [S\,{\sc ii}] image. In red is the integrated RHT intensity for parameters  $D_{\rm w}\,{=}\,15''$, $r_{\rm s}\,{=}\,2''$, and $Z\,{=}\,0.7$, overlaid on the [S\,{\sc ii}] image. The figure shows that the RHT is effective at recovering the straight radial filaments. 
Following \citet{clark2014-rht}, we adopt a threshold of $Z=0.7$. The remaining two parameters, $r_{\rm{s}}$ and $D_{\rm{w}}$, determine the spatial scales probed by the RHT and therefore depend on the characteristic size of the structures of interest. We explored a broad grid of parameter combinations, shown in Appendix Fig.\,\ref{fg:rht_scale5_grid}, to assess their effect on the recovered filament morphology. Smaller values of $r_{\rm{s}}$ and $D_{\rm{w}}$ are more sensitive to fine-scale structure and local variations in orientation, whereas larger values emphasize the global morphology and overall directionality of the filaments. In the next section, our goal is to characterize the large-scale organization and dominant orientation of the filamentary structures, and we therefore adopt comparatively large values of $r_{\rm{s}}$ and $D_{\rm{w}}$. In Section~\ref{sec:profile}, instead, we focus on the local properties of the filaments and consequently use smaller parameter values to better capture small-scale variations.

\subsection{Where is the nebula center?}

\noindent
A unique characteristic of the Pa\,30 nebula as a historical supernova of type Ia(x) is the presence of a central stellar remnant. 
Because the explosion in many type Iax models proceeds as a failed detonation or an off-center partial deflagration of the white dwarf, the surviving remnant can receive a substantial recoil kick \citep{jordan2012,lach2022}. Consistent with this picture, several high-velocity runaway stars ($>500$\,km\,s$^{-1}$) with atmospheres enriched by thermonuclear burning products have been identified as likely surviving remnants \citep{vennes2017,raddi2017,raddi2018a,raddi2018b,raddi2019,el-badry2023,bhat2026}. In contrast, other deflagration models predict only negligible remnant kicks \citep{kromer2013,fink2014}. 
Numerical predictions for the white dwarf kick range from 6 to 600\,km/s.

In the case of Pa\,30, the high-precision \textit{Gaia} proper motion of the central star corresponds to a transverse velocity of only $15.7\pm1.5$\,km\,s$^{-1}$ \citep{gaia2023}.
Measuring the radial velocity of the star is not feasible because its extreme wind velocity ($\approx16{,}000$,km,s$^{-1}$) produces emission lines that are tens of \AA\ wide and therefore unsuitable for precise radial-velocity measurements \citep{gvaramadze2019}. A line-of-sight kick component could instead be inferred from velocity asymmetries in the nebula. Using integral field unit spectroscopy, \citet{cunningham2024} report an intensity asymmetry of $\approx$40\% in [S\,{\sc ii}] emission, with a higher total intensity of emission from red-shifted material. However, they note that their observations cover only a small fraction ($\approx$10\%) of the nebula, and spectroscopic observations encompassing a larger area would be required to determine whether this asymmetry is a global property. Nevertheless, it is unlikely that any substantial kick velocity would be directed almost entirely along the line of sight, allowing us to rule out a large recoil kick. A modest kick, however, remains possible and would manifest itself in the plane of the sky as an offset between the current position of the central star and the geometric center of the nebula.
We therefore use the RHT applied to deep, high-resolution GMOS [S\,{\sc ii}] imaging to determine the sky position of the nebula center.

The RHT returns, for each pixel in the image, an angular intensity spectrum that quantifies the strength of linear structures as a function of orientation for angles ranging from 0--180\,degrees. In the lower panel of Fig.\,\ref{fg:rht_overlay}, we plot every pixel with some RHT intensity. The points are colored by the angle at which the pixel-specific RHT spectrum peaks, with a color map that is cyclical and spans 0--180 degrees. The figure demonstrates that the RHT is sensitive not just to the presence of linearity, but also that the radial angle is well recovered. 
This behavior reflects the fact that the filaments possess a strong directional component, with their spatial extent preferentially aligned along the radial direction. The RHT therefore provides a quantitative confirmation of the visually apparent radial morphology of the filaments.

\begin{figure}
    \centering 
    \includegraphics[width=.9\columnwidth]{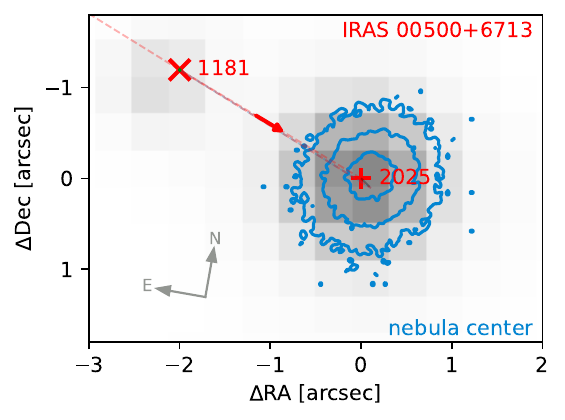}\\[-1ex]
    \includegraphics[width=.96\columnwidth]{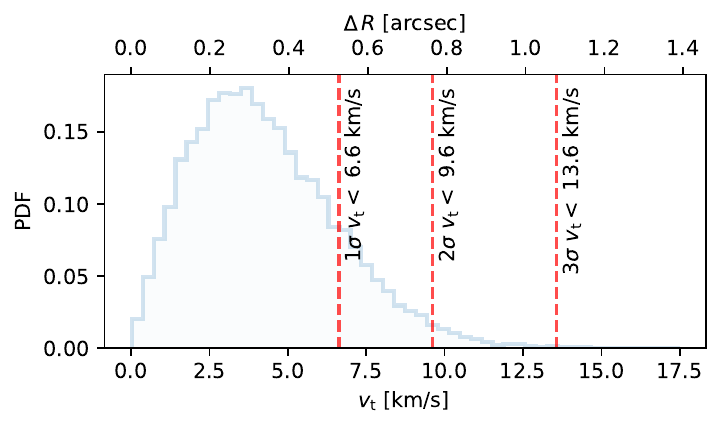}\\[-1ex]
    \includegraphics[width=1.0\columnwidth]{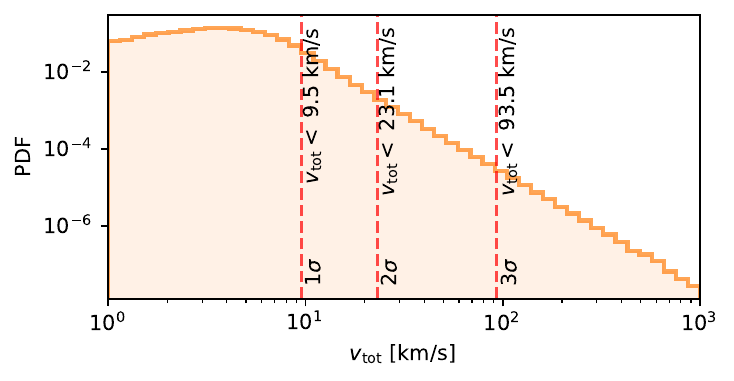}
    \caption{\textbf{Top}: In grey, we show a cutout of the same [S\,{\sc ii}] GMOS image from Fig.\,\ref{fg:ql_4panel_south}, focused within a few arcsec of the central star. In blue, we show the measured center of the nebula from our RHT MCMC analysis, with 1, 2, and 3-sigma uncertainty contours. The position of the central star, IRAS\,00500+6713, at the 1181 and 2025 epochs, is shown in red. We find that the nebula center and central star show a common sky position, ruling out separations larger than 1.1'' at 3\,$\sigma$.  \textbf{Middle}: Distribution of transverse velocities consistent with the sky separation of the measured nebula center and the position of the central star. These analysis support a 3\,$\sigma$ upper limit on the transverse velocity of $v_{\rm t}\,{<}\,14$\,km/s. \textbf{Bottom}: The probability density function (PDF) for the total kick velocity, computed by propagating the posterior on transverse velocity (middle panel) under the assumption of equal probability for all kick trajectories in three-dimensional space.}
 	\label{fg:rht_kick}
\end{figure}

Assuming that the filaments collectively point back to a common nebula center, we now develop a technique to measure the center and compare it to the location of the central star. As we are interested in the overall directionality of the filaments, and not in their fine structure, we smooth and resample the [S\,{\sc ii}] image to 0.4'', slightly below the 0.5'' seeing. This reduces the number of pixels and increases the signal-to-noise per pixel. To quantify the filament orientations, we first compute, for every pixel, the total RHT intensity summed over all angles. This quantity measures how strongly a given pixel participates in any linear structure. To further suppress noise, we discard the 10\% of pixels with the lowest intensity. For each of the retained pixels, we define a single characteristic orientation by taking the angle at which the RHT intensity spectrum reaches its maximum. This produces a map of local filament directions across the nebula (similar to what is shown in the lower panel of Fig.\,\ref{fg:rht_overlay}).

We then construct a simple geometric model in which the filaments are assumed to be radial with respect to a common center $(x_c,y_c)$. For a given trial center, we calculate at each selected pixel the angle of the radial line connecting the trial center to that pixel. These predicted angles are then compared with the corresponding RHT-derived orientations. We quantify the goodness of fit using the mean absolute angular difference between the predicted and measured angles. Trial centers that produce smaller angular residuals correspond to stronger radial coherence and are therefore assigned higher likelihoods. We also include, as a free parameter, the intrinsic scatter in the angular resolution, which allows us to model the noise in the data and procedure.

We restrict the trial centers to lie within a box of $\pm15''$ centered on the position of the central star, adopting a uniform prior within this region and zero probability outside it. We sample the posterior distribution using the Markov Chain Monte Carlo ensemble sampler implemented in \texttt{emcee} \citep{mackey2013-emcee} (see Appendix\,\ref{sec:appendix-model} for details). 

To test whether the choice of parameters for the RHT influences the results of our fit, we perform the same fit for a grid of different $D_{\rm{w}}$ and $r_{\rm{s}}$. The best-fit value for the center of the nebula does not depend on the choice of parameters, but less appropriate choices yield larger uncertainties. The full model description and posterior distributions are presented in Appendix Section~\ref{sec:appendix-model}.

Our best fit for the nebula center is shown in Fig.\,\ref{fg:rht_kick}. The upper panel shows the original [S\,{\sc ii}] image, zoomed into a $\approx$3 arcsec box around the central star, IRAS\,00500+6713. 
The red addition symbol denotes the Gaia position of the central star at the epoch 2025.6 of the GMOS observations. We find this is in good agreement with the source visible in the image. The red cross shows the expected sky position of the central star in the year 1181, derived using the Gaia proper motion.\footnote{The proper-motion-corrected position of IRAS\,00500+6713 for the year 1181 happens to fall very close to another source in the field, Gaia DR3 526287189166341504. The spatial association is coincidental. Gaia DR3 526287189166341504 has proper motion of $\sim$10\,mas per year, and a Gaia distance of $D_1 \approx 988 \pm 216\ \text{pc}$.}
Overlaid in blue are the one, two, and three sigma contours of the fitted nebula center. These contours express the mixed posterior resulting from fits with a range of RHT parameters ($D_{\rm w}\,{=}\,$12'', 13'', 15'', 17'', $r_{\rm s}\,{=}\,$2'' and 2.4'', and $Z\,{=}\,0.7$), allowing us to marginalize over the choice of RHT parameters (see Appendix Fig.\,\ref{fg:rht_scale5_grid} for the individual posteriors from this grid). We can see that the derived center is fully consistent with the position of the central star, indicating that if the star received a kick in the explosion, it must be small. From the posteriors of our fit, we can derive a limit on the transverse velocity of the kick in the plane of the sky as $v_{\rm t}\,{<}\,14$\,km/s at $3\sigma$.

We propagate the posterior in transverse velocity into limits on the total three-dimensional kick velocity by adopting the assumption that in 3D space all kick directions are equally likely. The lower panel of Fig.\,\ref{fg:rht_kick} shows the probability density function (PDF) for the total kick velocity, which yields an upper limit of $v_{\rm tot}\,{<}\,94$\,km/s at $3\sigma$.

\subsection{Characteristic filament profile}
\label{sec:profile}

\begin{figure*}
	\centering    
\includegraphics[width=1.02\columnwidth]{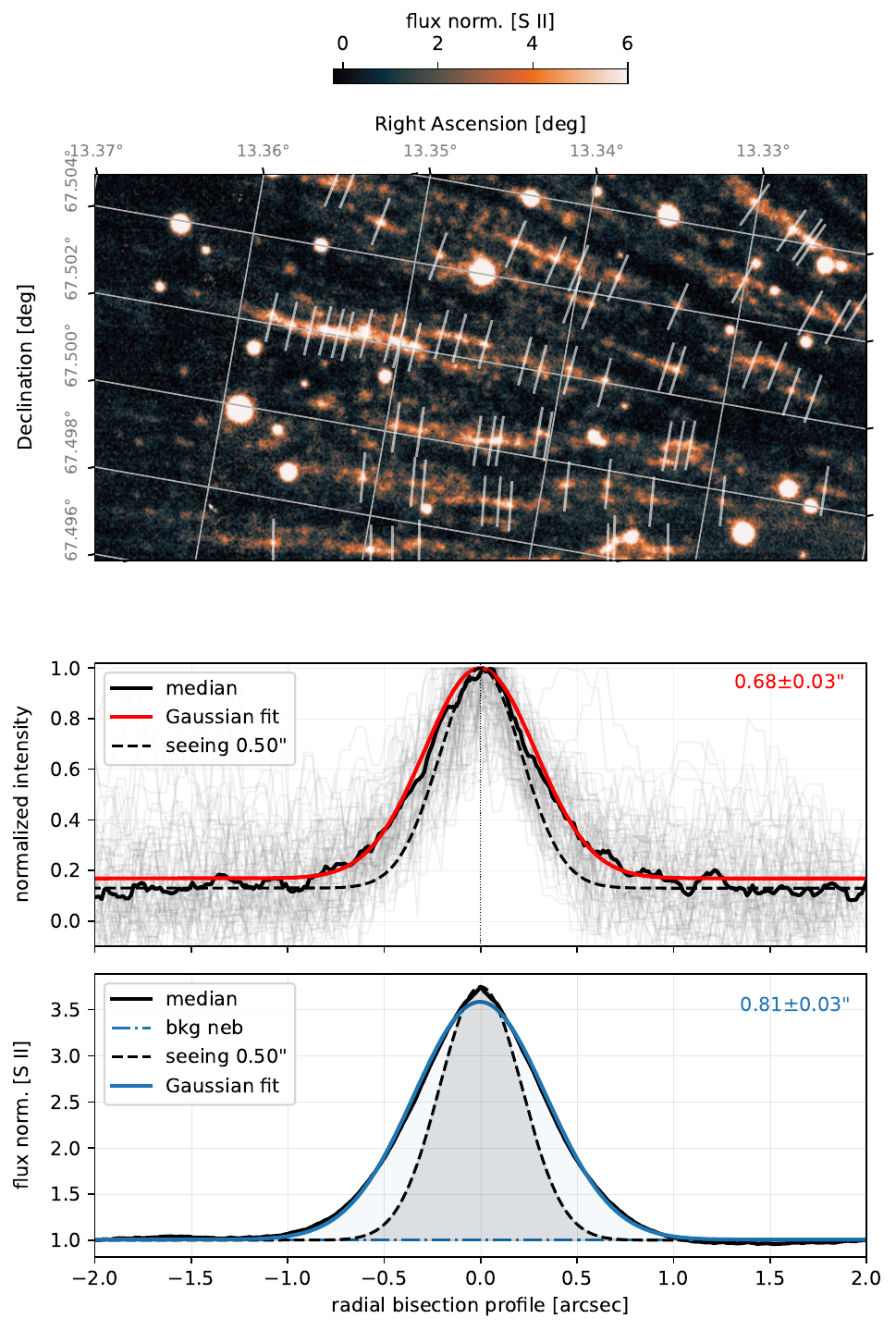} 
\includegraphics[width=1.02\columnwidth]{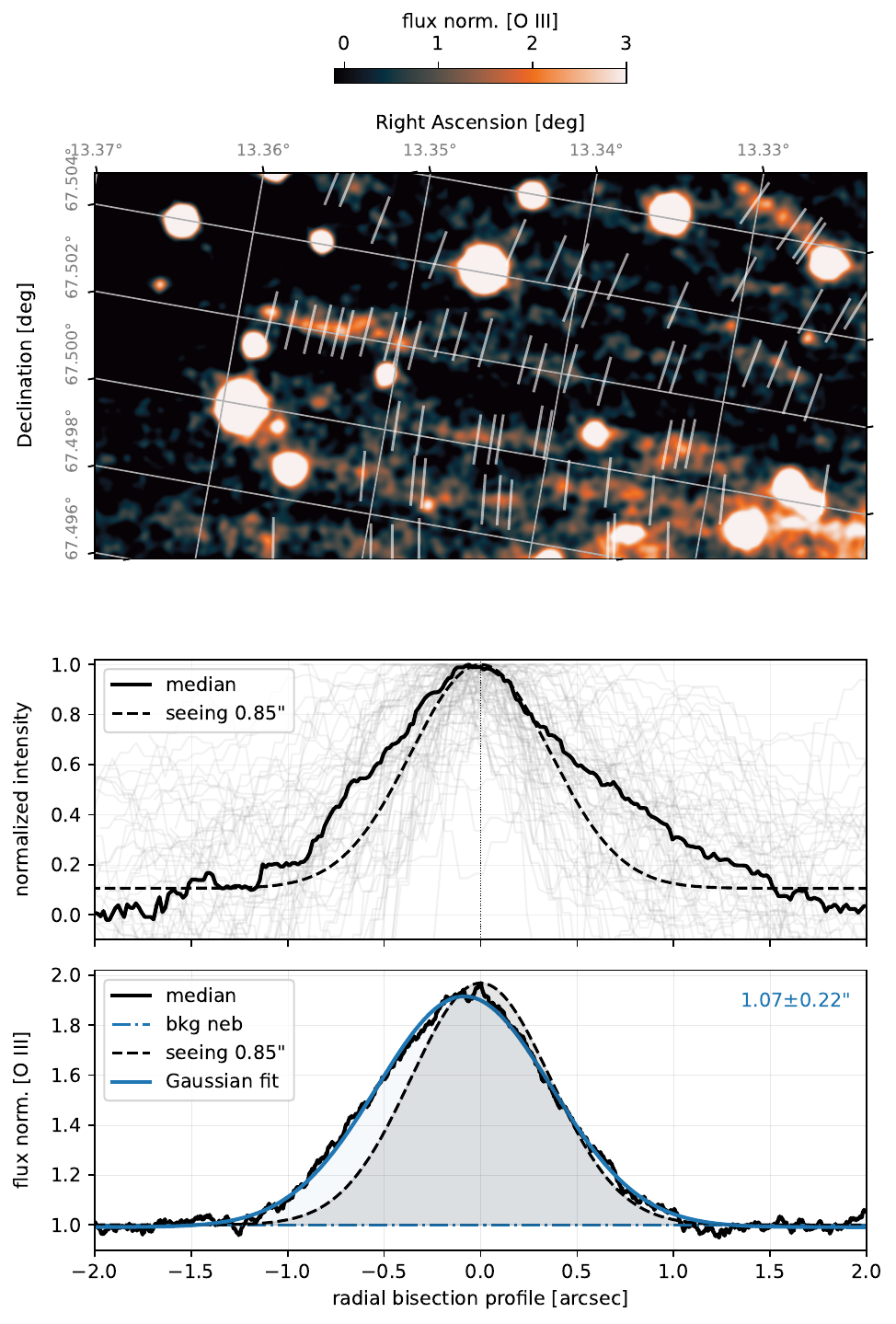} 
\caption{\textbf{Left}: The top panel shows the background-subtracted [S\,{\sc ii}] image, focused on the East edge of the nebula. The image has been normalized by the ``nebula background''. The middle panel shows in grey the perpendicular profile for dozens of 4-arcsec cross-sections placed across visible filaments and knots. The perpendicular direction is defined with respect to the nebula center. The locations and orientations of these profiles are shown in the upper panel with the short grey lines. The flux in all profiles is background-subtracted and normalized to the peak. We measure a filament FWHM by fitting a Gaussian (in red) to the median profile (in black). We find a median filament cross-section of FWHM$\,=\,$0.68$\,\pm\,$0.02'', larger than the empirical PSF (0.50$\,\pm\,$0.01''). In the lower panel, we show the background-subtracted median profile for all RHT filaments found in the [S\,{\sc ii}] GMOS imaging (in black). The profile has been normalized by the ``nebula background''. The nebula background, indicated with the dot-dash horizontal line, is an empirical detection of an additional ``background'' between filaments which is in excess over the sky background (see Fig.\,\ref{fg:rht_profiles_annuli_2panel}). We again fit a Gaussian (in blue) to the profile and recover a FWHM$\,=\,$0.81$\,\pm\,$0.03''.
\textbf{Right}: Similar to the left panels, but for [O\,{\sc iii}]. We omit a fit to the image filaments (middle panel) but fit the full median RHT profile in the lower panel, finding a FWHM$\,=\,$1.07$\,\pm\,$0.22''.}
	\label{fg:img_profiles_bespoke}
\end{figure*}

\noindent
Fig.\,\ref{fg:ql_4panel_south} shows that the filaments are composed of tight chains of knots, whose apparent sizes remain approximately constant across the nebula. To determine whether these knots are resolved in the image and to measure their characteristic size, we extract flux profiles across several bright filaments. For each filament, we define a cross-sectional profile perpendicular to the local radial direction, where the radial axis is defined relative to the common nebula center derived in the previous section. Each extracted profile spans a length of $4''$, centered on each filament.
The upper panels of Fig.~\ref{fg:img_profiles_bespoke} show the selected filaments in the eastern region of the nebula, together with the orientation and extent of the extracted cross-sections in both the [S\,{\sc ii}] and [O\,{\sc iii}] images. The middle panels display the corresponding flux profiles for each cross-section, along with the median profile. All profiles have been sky-subtracted and normalized to their peak flux. For [S\,{\sc ii}], a Gaussian fit to the median profile yields a FWHM of 0.68\,$\pm$\,0.03''; given the seeing of 0.5\,$\pm$\,0.01'' (see section~\ref{sec:obs}), this measurement provides evidence that the knots in the [S\,{\sc ii}] imaging are marginally resolved. Although the images have been sky-subtracted, the [S\,{\sc ii}] profiles do not fall to zero outside the filament cores. Instead, the emission levels off to a local background component that remains significantly above the sky level, suggesting the presence of diffuse emission surrounding the knots. For the [O\,{\sc iii}] image, the extracted profiles are too noisy to get a reliable fit, unless some smoothing is applied to the image, reducing the resolution. In the middle right panel of Fig.~\ref{fg:img_profiles_bespoke}, we show the profiles extracted after smoothing the image to half of the seeing. However, because of the poorer image quality, we do not expect to resolve the filaments in the  [O\,{\sc iii}] image.

\begin{figure}
	\centering    
\includegraphics[width=1.0\columnwidth]{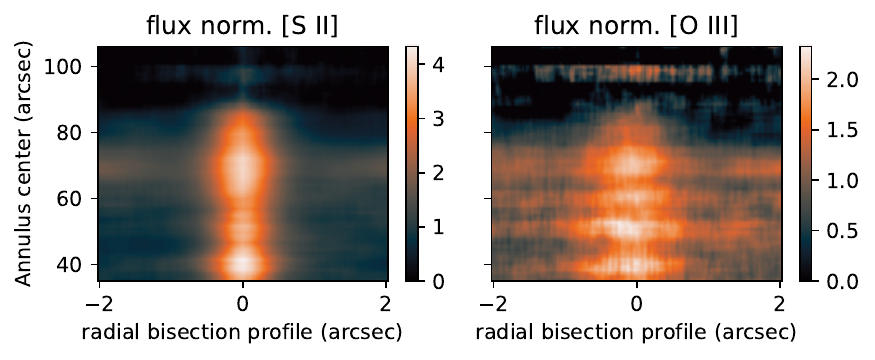} 
\caption{\textbf{Left}: Median perpendicular profiles, similar to the lower panels of Fig.\,\ref{fg:img_profiles_bespoke}, but as a function of distance from the nebula center. The presence of the nebular background is clearer here and can be seen to reach the level of the sky background at around 90'' from the nebula center. 
\textbf{Right}: Similar to the left panel, but for [O\,{\sc iii}].}
	\label{fg:rht_profiles_annuli_2panel}
\end{figure}

For this analysis, we manually picked a small number of filaments (approximately 30) across an arbitrary region of the nebula.
However, we have shown that the RHT intensity provides a sensitive and robust tracer of the spatial location of filamentary structures, allowing us to extend this analysis across the entire nebula in a systematic way. Rather than selecting filaments by eye, we extract a perpendicular flux profile at every pixel with significant RHT intensity. As before, each profile is taken perpendicular to the radial direction defined by the line connecting the pixel to the derived center of the nebula, extending $2''$ on either side of the RHT-bright pixel. This approach yields thousands of profiles distributed across the full image.
For both the [S\,{\sc ii}] and the [O\,{\sc iii}] images, we use the pixel locations extracted from the RHT map of the [S\,{\sc ii}] image. The surface brightness of the [S\,{\sc ii}] image is significantly higher, producing a more reliable and better-defined distribution of significant RHT intensity.
The bottom panels of Fig.\,\ref{fg:img_profiles_bespoke} show the median perpendicular profiles derived across the nebula for both the [S\,{\sc ii}] and [O\,{\sc iii}] images, at native resolution (pixel scale of 0.08''). The flux values here have been sky-subtracted and then normalized to the value of the ``local background''. As already suggested by the manually extracted profiles, the emission does not decline directly to the sky level outside the filament cores. Instead, the profiles flatten to a diffuse local background component that is brighter than the sky. This behavior is shown more clearly in Fig.~\ref{fg:rht_profiles_annuli_2panel}, where the nebular background decreases toward the sky level only at a distance greater than $\approx90''$ from the center of the nebula. A Gaussian fit to the median profile yields a FWHM of 0.81\,$\pm$\,0.03'' for [S\,{\sc ii}] and of 1.07\,$\pm$\,0.22'' for [O\,{\sc iii}]. Again, the size of the [S\,{\sc ii}] filaments appears marginally resolved compared to the seeing, while the FWHM of the [O\,{\sc iii}] profile is consistent with the poorer image quality of the oxygen image (with a median seeing of 0.85\,$\pm$\,0.02''). The [O\,{\sc iii}] plot shows a slight shift in the filament profile with respect to the centered Gaussian fit; however, the shift is likely not significant due to the low signal-to-noise in the oxygen data, and deeper imaging is required to confirm it.

Several features emerge from this RHT perpendicular profile analysis. 
First,  the median profiles derived from the full RHT-selected sample closely resemble those obtained from the manually-selected bright filaments, while exhibiting substantially reduced noise owing to the much larger number of measurements.
Second, we confirm that the filament knots are marginally resolved in the [S\,{\sc ii}] image with a characteristic FWHM of order $0.7-0.8''$. De-convolving the PSF at the [S\,{\sc ii}] image seeing, we derive a typical angular size of $\approx$0.6'', or ${\sim}10^{16}$~cm at the distance of the nebula \citep[$D=2.3\pm0.1$\,kpc,][]{gaia2021}.
Third, we find that the wings of the profiles for both [S\,{\sc ii}] and [O\,{\sc iii}] approach a constant flux level above the sky background, defining an effective ``nebular background'' component. While the nebula background could originate from nearby resolved filaments that effectively contaminate the wings of the profiles, we do not consider this the likely explanation since the characteristic profile is derived via the median statistic, which should be robust to contaminants (e.g., nearby stars and filaments). Inspection of the images in Fig.\,\ref{fg:img_profiles_bespoke} shows that the typical cross-sectional profiles do not intersect additional resolved filaments in the majority of instances. Thus, the nebula background likely originates from one of two possibilities. It may reflect the presence of a diffuse emission filling the regions between the bright filaments, or it may arise from a large population of faint filaments that collectively appear diffuse.

\section{Summary} 
\label{sec:summary}
We have presented deep, narrowband Gemini North GMOS imaging of the Pa\,30 nebula that reveals remarkably rich small‑scale structure. The [S\,{\sc ii}] image shows that the number of resolvable filaments in the nebula is an order of magnitude larger than previously inferred from narrowband imaging \citep{fesen2023}. Furthermore, inspection of individual filaments reveals that, in general, a filament is not a continuous radial structure, but rather a sequence of knots or clumps with some characteristic size and separation. \change{As the [S\,{\sc ii}] filaments are marginally resolved, we can measure a characteristic knot diameter of approximately 0.6'', corresponding to ${\sim}10^{16}$\,cm. The separation between knots may also have a characteristic scale, as many filaments show a regular spacing of their constituent knots. Quantifying this scale directly, however, is complicated by projection effects: individual filaments can overlap along the line of sight, making it difficult to distinguish the spacing of knots within a single filament from apparent separations produced by superposed structures. A robust empirical measurement of the characteristic knot separation is therefore non-trivial. Nevertheless, models that predict a characteristic spacing could be tested through forward modeling, by projecting synthetic filament populations and comparing their observed knot distributions directly with the data.} Complementary [O\,{\sc iii}] imaging demonstrates that the nebula is also threaded by filaments in this higher‑ionization line; \change{as we show in Appendix Section~\ref{sec:cross-correlation},} these [O\,{\sc iii}] filaments are co-spatial with the [S\,{\sc ii}] structures.

To quantify the ``radiality'' of filaments in the nebula, we apply the Rolling Hough Transform (RHT) to the [S\,{\sc ii}] data. The RHT recovers the filamentary network and allows us to assign, for each pixel, a preferred orientation angle. By comparing these RHT‑derived orientations to the directions of rays drawn from trial centers on the sky, we show that the filaments are collectively biased toward a radial geometry. Using an MCMC exploration of the parameter space of possible centers, we identify the point on the sky that best reproduces the observed distribution of filament orientations. The resulting nebular center is consistent, within $\approx$1'', with the \textit{Gaia} position of the central star. This close correspondence implies that the stellar remnant has not experienced any substantial kick in the plane of the sky relative to the nebula.

The RHT map also provides a convenient way to characterize the transverse structure of the filaments. By extracting thousands of perpendicular cuts through the [S\,{\sc ii}]‑selected filament pixels, we construct median cross‑sectional profiles in both [S\,{\sc ii}] and [O\,{\sc iii}]. These profiles exhibit a pronounced central excess superposed on a “nebular background” level that lies above the sky, indicating either genuinely diffuse emission between the filaments or a network of unresolved filaments fainter than our detection threshold. From the median profiles, we measure a characteristic full width at half maximum for the filaments and derive a representative perpendicular intensity profile for [S\,{\sc ii}] and [O\,{\sc iii}], which suggests that the filaments are marginally resolved in the [S\,{\sc ii}] imaging that was obtained with 0.50'' seeing. In contrast, the [O\,{\sc iii}] data, obtained under less favorable seeing ($\approx$0.85''), do not show a clear excess over the instrumental PSF, and thus do not independently require the filaments to be spatially resolved.

As discussed in section~\ref{sec:knots}, the discovery and characterization of the size of knots in the filaments provide important constraints on models proposed to explain the peculiar morphology of Pa\,30. Quantitative constraints on the recoil kick of the central star can provide hints on the progenitor of the type Iax supernova. Further observations exploiting the exquisite angular resolution of the Hubble Space Telescope would greatly enhance this analysis.

\section*{Acknowledgements}
\noindent
\change{We thank Jim Fuller, SR Kulkarni, David Charbonneau, Paul Duffell, Abigail Polin, Miranda Pikus, Tony Piro, Lars Bildsten, Eliot Quartert, and Rob Fesen for helpful discussions.}
This work was supported by NASA through the NASA Hubble Fellowship grant HST-HF2-51527.001-A awarded by the Space Telescope Science Institute, which is operated by the Association of Universities for Research in Astronomy, Inc., for NASA, under contract NAS5-26555.


%

\vspace{5mm}
\facilities{Gemini North: GMOS-N}


\software{\textsc{astropy} \citep{2013A&A...558A..33A,2018AJ....156..123A},  
\textsc{scipy} \citep{scipy2020}, 
\textsc{emcee} \citep{mackey2013-emcee}, \textsc{DRAGONS} \citep{labrie2023-dragons}, \textsc{DS9} \citep{joye2003-ds9}}





\appendix


\section{The Rolling Hough Transform}
\label{sec:RHTexpl}

\begin{figure}
	\centering    
\includegraphics[width=1.0\columnwidth]{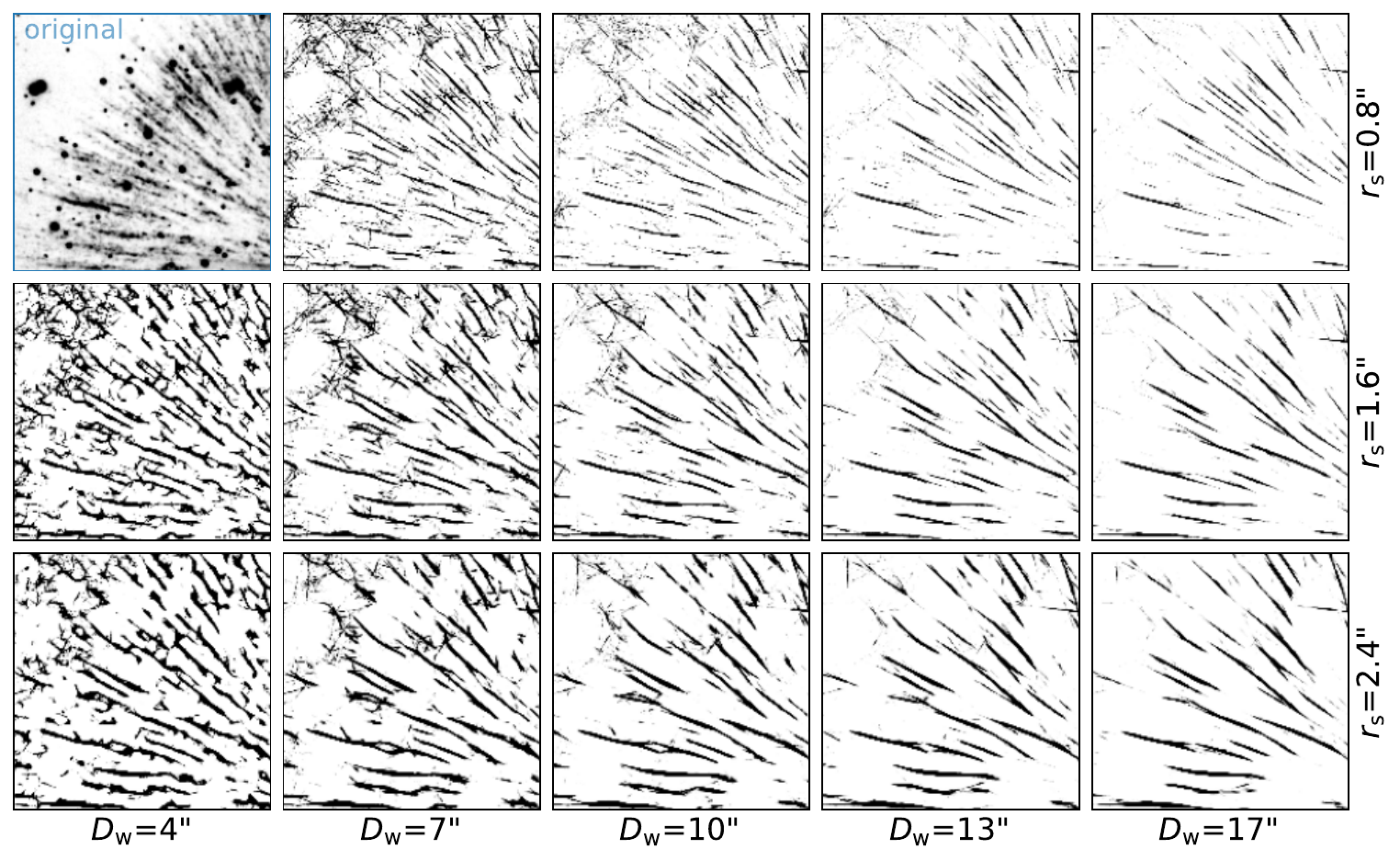} 
\caption{The top-left panel shows the North-East quadrant from the [S\,{\sc ii}] image shown in the top-left panel of Fig.\,\ref{fg:ql_4panel_south}. The other panels depict a grid of RHTs for a range of parameters. Defined in units of pixels for the down-sampled image (pixel scale 5\,$\times$\,0.08\,$=$\,0.4''), from left-to-right, the RHT increases in window length ($D_{\rm w}$), and from top-to-bottom in the smoothing radius ($r_{\rm s}$).}
	\label{fg:rht_trial_grid}
\end{figure}

The optimal choice of parameters for the RHT depends on the characteristic spatial scales of the structures one aims to recover. To assess the most appropriate parameter choices for our analysis, we generated RHT maps over a broad grid of window lengths ($D_{\rm w}$) and smoothing radii ($r_{\rm s}$). In Figure~\ref{fg:rht_trial_grid} shows a representative subset of this grid, focused on the north-eastern quadrant of the [S\,{\sc ii}] image. The window length $D_{\rm w}$ primarily controls the spatial scale over which filament orientations are measured. Smaller values of $D_{\rm w}$ make the RHT more sensitive to fine-scale structure and local variations in filament direction, but also increase the level of noise in the recovered map. In contrast, larger values emphasize the global morphology and large-scale coherence of the filaments, producing smoother and more continuous structures. The smoothing radius $r_{\rm s}$ determines the characteristic transverse scale of the structures retained by the RHT. When $r_{\rm s}$ becomes comparable to the filament width, as in the case of $r_{\rm s}=0.4$'', portions of the filamentary structure begin to be suppressed or washed out. Conversely, excessively large values of $r_{\rm s}$ cause neighboring filaments to blend together, reducing the contrast between adjacent structures.

In Section~\ref{sec:rht}, our primary goal is to recover the dominant large-scale orientation of the filamentary structures. We therefore adopt relatively large values of the window length, $D_{\rm w}$, together with modestly large smoothing radii, $r_{\rm s}$, to emphasize coherent global morphology while suppressing small-scale fluctuations. Specifically, we use $D_{\rm w} = 12$, 13, 15, and 17'', together with $r_{\rm s} = 2$ and 2.4'' (see next section).

In contrast, the analysis presented in Section~\ref{sec:profile} focuses on the fine-scale internal structure of the filaments, including the knotty substructures. For this purpose, we perform this analysis on the original GMOS image in its native pixel scale of 0.08'', adopting smaller parameter values that preserve local variations and enhance sensitivity to compact features, using $r_{\rm s}=0.8$'' and $D_{\rm w}=3.3$''.

\begin{figure}[h]
	\centering    
\includegraphics[width=1.0\columnwidth]{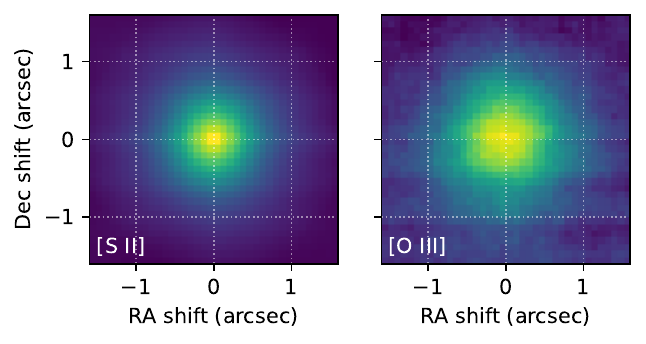} 
\caption{\change{Spatial cross-correlation of the narrowband imaging against the output of the [S\,{\sc ii}] RHT. The left panel serves as a control and shows the cross-correlation of the [S\,{\sc ii}] image and the [S\,{\sc ii}] RHT output. The right panel shows the cross-correlation of the [O\,{\sc iii}] image and the [S\,{\sc ii}] RHT output. This analysis provides strong evidence that the [O\,{\sc iii}] filaments are co-located with those emitting in [S\,{\sc ii}].}}
	\label{fg:cross-correlation-OIII-SII}
\end{figure}

\change{
\section{Spatial correlation of [S\,{\sc ii}] and [O\,{\sc iii}] filaments}
\label{sec:cross-correlation}
We investigate the spatial correlation between the [S\,{\sc ii}] and [O\,{\sc iii}] nebula emission by performing a two-dimensional (2D) cross-correlation of the [S\,{\sc ii}] Rolling Hough Transform against the [O\,{\sc iii}] image. The results of the 2D cross-correlation are shown in the right panel of Fig.\,\ref{fg:cross-correlation-OIII-SII}. The cross-correlation is performed on normalized images, and intensity values exceeding those of brightest filaments are masked -- effectively masking all stars. Intensity values lower than the faintest filaments are also masked; effectively masking background-dominated regions. 
}

\change{
We compute the correlation amplitude, $C$, between two normalized images as
\begin{equation}
    C = \frac{\sum\limits_{n=0}^{N}(I_{i}^n - \bar{I_{i}})  (I_{j}^n - \bar{I_{j}})}
    {\left(\sum\limits_{n=0}^{N}(I_{i}^n - \bar{I_{i}})^2  \sum\limits_{n=0}^{N}(I_{j}^n - \bar{I_{j}})^2\right)^{1/2}}
\end{equation}
where $I_{i}^n$ is the intensity of a given pixel ($n$) in either image $i$ or $j$, and $N$ is the total number of overlapping pixels. This amplitude is computed for integer pixel shifts in all directions across the plane of the sky. We perform this procedure for the [O\,{\sc iii}] image against the output of the RHT analysis on the [S\,{\sc ii}] image. As a control, we also perform the analysis on the [S\,{\sc ii}] image against its own RHT output. This control should, by definition, show higher spatial correlation toward zero shifts. This behavior is shown in the left panel of Fig.\,\ref{fg:cross-correlation-OIII-SII}. We find very similar behavior in the [O\,{\sc iii}] cross correlation (right panel), demonstrating that the [O\,{\sc iii}] filaments are highly correlated with the [S\,{\sc ii}] filaments, with spatial correlation increasing towards smaller shifts.
}

\section{Statistical model for finding nebular center}
\label{sec:appendix-model}

\noindent
At the native resolution of the [S\,{\sc ii}] image, the RHT identifies approximately two million pixels with intensities above the significance threshold (Z).
To reduce the data volume and increase the signal-to-noise ratio per pixel, we smooth the image and resample it to 0.4'', slightly below the seeing, \change{reducing the number of significant RHT pixels to approximately 30,000. Because the filaments close to the central star are predominantly oriented along the line of sight \citep{cunningham2024}, and therefore show weaker projected directionality than those at larger radii, we restrict the analysis to an annulus between 30'' and 90'' (close to the outer edge of the nebula).} Finally, to suppress noise, we discard the 10\% pixels with the lowest RHT intensity. \change{This selection emphasizes strong, well-defined filaments while maintaining} approximately uniform spatial coverage.
\change{For each retained pixel (i), the RHT provides its image coordinates $(j_i,i_i)$, its preferred orientation $(\theta_i)$, and its maximum RHT intensity $w_i>0$, which we use as a weight. The resulting RHT orientations are illustrated in Fig.\,\ref{fg:rht_overlay}.}

\change{We determine the nebular center by finding the point about which the observed filament orientations are most nearly radial.} The model has three free parameters: the coordinates of the trial center in pixel units, $\mathbf{c} = (x_c, y_c)$, and an intrinsic angular scatter in the RHT intensity spectra in degrees, $\sigma_0 > 0$. For sampling, we parameterize the latter as $\eta = \log \sigma_0$.
For a trial center $\mathbf{c}$, the expected radial orientation at pixel $i$ is
\begin{equation}
  \phi_i(\mathbf{c})
  =
  \left[
    M - \frac{180}{\pi}
    \arctantwo\;\!\bigl(i_i - y_c,\; j_i - x_c\bigr)
  \right] \bmod M ,
\end{equation}
where, $M=180^\circ$. Orientations are defined modulo $180^\circ$, since a filament at angle $\theta$ is indistinguishable from one at $\theta+180^\circ$.
The wrapped angular residual between the observed and expected orientations is
\begin{equation}
  \Delta\theta_i
  =
  \bigl(\theta_i - \phi_i(\mathbf{c}) + \tfrac{M}{2}\bigr) \bmod M
  - \tfrac{M}{2},
\end{equation}
so that the maximum angular residual is 90 degrees.
\change{We assume that these residuals are independent and Gaussian, with their variance inversely weighted by the RHT intensity ($\sigma_0^2/w_{\rm i}$).}
Thus, pixels with stronger RHT detections contribute more strongly to the fit. The corresponding log-likelihood is
\begin{equation}
  \log \mathcal{L}(\mathbf{c}, \sigma_0)
  =
  -\frac{1}{2}
  \sum_{i=1}^N
  \left[
    \frac{\Delta\theta_i^2}{\sigma_0^{2} / w_i}
    +
    \log\!\left(
      2\pi\,\frac{\sigma_0^{2}}{w_i}
    \right)
  \right].
  \label{eq:loglike}
\end{equation}
The quadratic (first) term rewards small residuals, whilst the logarithmic (second) term penalizes overly broad scatter.
\change{We adopt a uniform prior on the center within $\pm16''$ ($\pm200$ pixels) of a reference position $(x_{c0},y_{c0})$ in both coordinates, and a log-uniform prior on $\sigma_0$ between $0.5^\circ$ and $60^\circ$. Defining the sampled parameter vector as $\Theta=(x_c,y_c,\eta)$, where $\eta=\log\sigma_0$, the posterior is}
\begin{equation}
  \log p(\Theta \mid \{\theta_i, j_i, i_i, w_i\})
  =
  \log p(\mathbf{c})
  +
  \log p(\eta)
  +
  \log \mathcal{L}(\mathbf{c}, \sigma_0),
\end{equation}
The first term, $\log p(\mathbf{c})$, denotes the uniform prior on the nebula center, the second term, $\log p(\eta)$, denotes the prior on $\eta = \log\sigma_0$ (uniform in $\eta$ between $\log 0.5$ and $\log 30$), and the third term, $\log \mathcal{L}(\mathbf{c}, \sigma_0)$, denotes the Gaussian log-likelihood of all the wrapped angular residuals.
\begin{figure}[t]
	\centering    
\includegraphics[width=1.0\columnwidth]{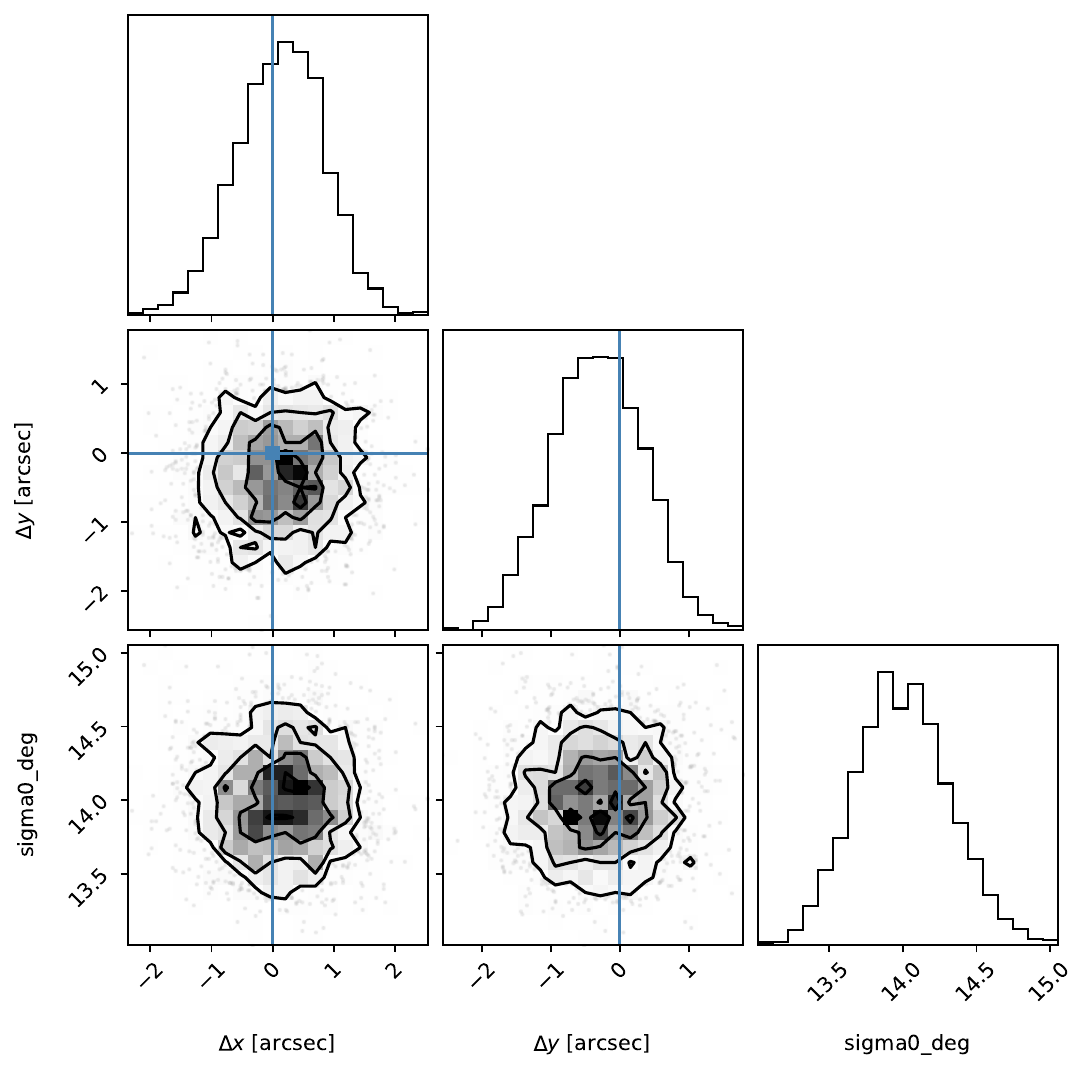} 
\caption{The posterior distributions for the three fitted parameters, after the disposal of the ``burn-in'' steps. This figure is for one specific choice of RHT parameter combination ($D_{\rm w}\,{=}\,15$'', $r_{\rm s}\,{=}\,2$'', and $Z\,{=}\,0.7$).}
	\label{fg:rht_find_center_chains_corner}
\end{figure}

\change{We sample the posterior using \texttt{emcee} \citep{mackey2013-emcee}, adopting 24 walkers evolved for 1000 steps. The walkers are initialized around the \textit{Gaia} position of IRAS,00500+6713 with a dispersion of 2 pixels in each coordinate, and around $\sigma_0=5^\circ$, with a dispersion of $2^\circ$. We discard the first five autocorrelation times, corresponding to approximately 200 steps, as burn-in. Figure,\ref{fg:rht_find_center_chains_corner} shows an example of the resulting posterior distributions, which are approximately Gaussian and show little covariance between the fitted parameters.}

\begin{figure*}[h]
	\centering    
\includegraphics[width=0.8\textwidth]{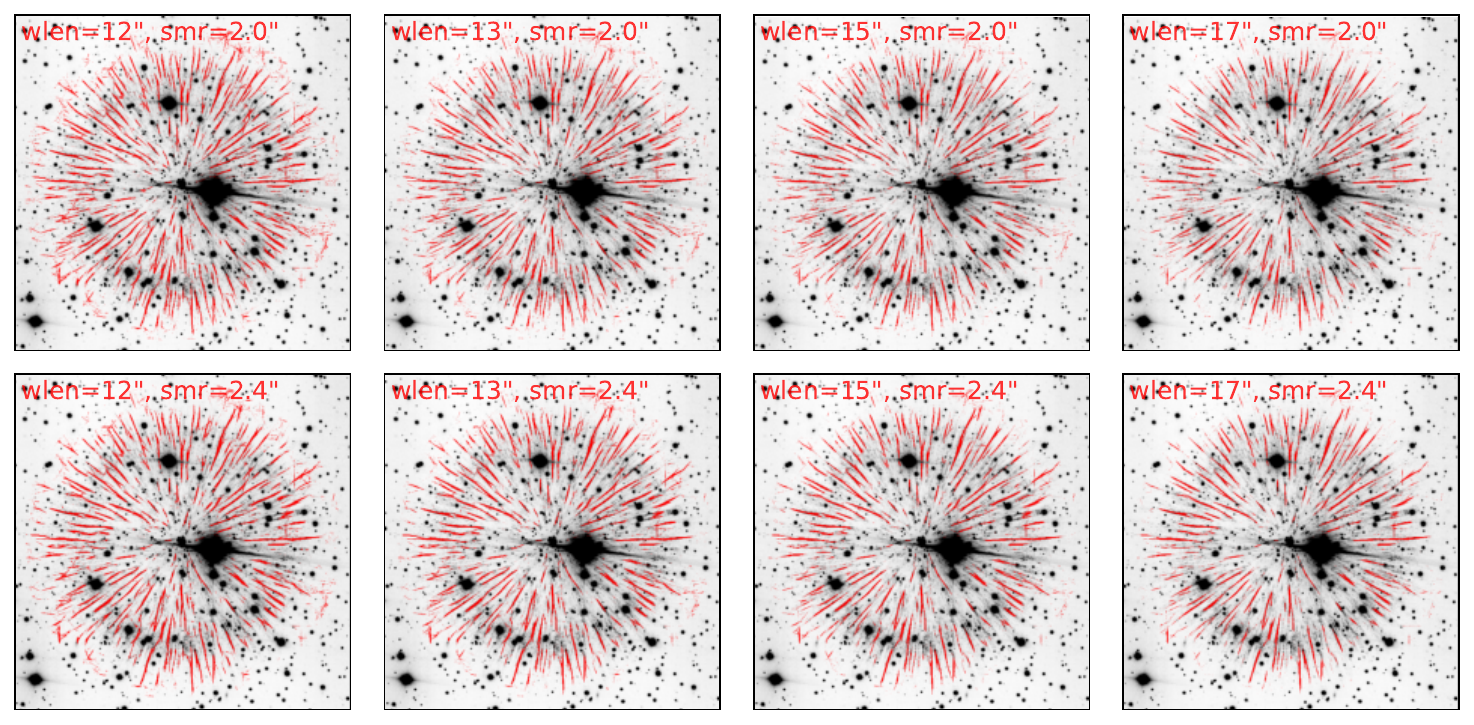}  \\
\includegraphics[width=0.88\textwidth]{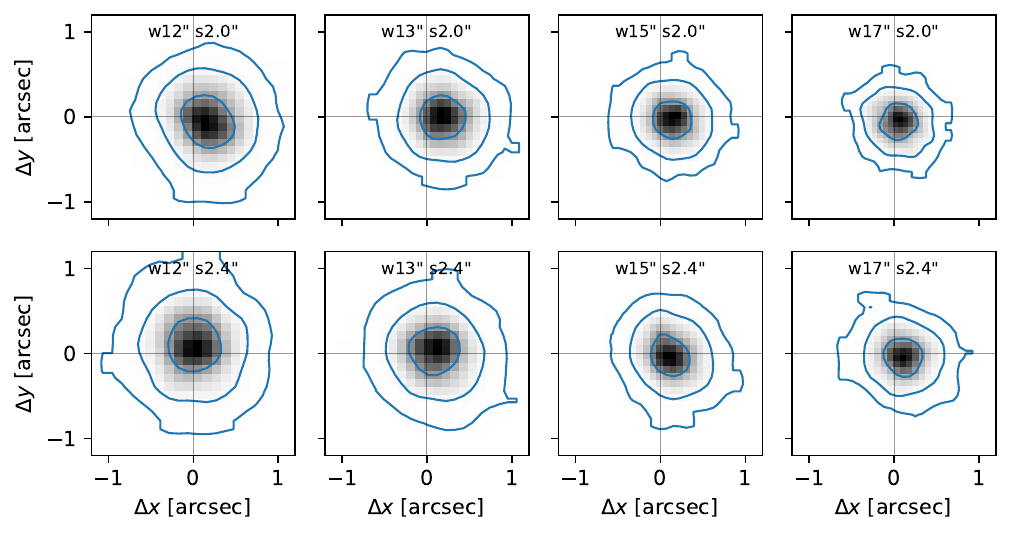} \hspace{30pt}
\caption{\change{Results of our analysis to determine the nebular center for different choices of RHT parameters. The upper panels show the integrated RHT intensity for every pixel in red overlaid over the [S\,{\sc ii}] image, while the lower panels show the resulting posteriors for the coordinates of the nebular center. The upper and lower rows show the results for $r_{\rm{s}}\,{=}\,$2'' and 2.4'' respectively, while the columns, from left to right, are for $D_{\rm{w}}\,{=}\,$12'', 13'', 15'', 17''.}}
	\label{fg:rht_scale5_grid}
\end{figure*}

To assess the sensitivity of our results to the choice of RHT parameters, we repeat the analysis over a broad grid of window lengths, $D_{\rm w}$, and smoothing radii, $r_{\rm s}$. We find that the best-fitting values for $\Delta x$ and $\Delta y$ remain consistent across the full range of parameter combinations explored, indicating that the inferred nebula center is robust against the specific RHT parameter choice. However, for suboptimal parameter combinations, the posterior distributions become significantly broader, leading to larger uncertainties. The most stable and tightly constrained solutions are obtained for $D_{\rm{w}}\sim10$''$-20$'' and $r_{\rm{s}}\sim2$''$-3$''. To marginalize over our choice of parameters by combining the posterior distributions from the subset of runs shown in Fig.\,\ref{fg:rht_trial_grid}, corresponding to $D_{\rm{w}}\,{=}\,$12'', 13'', 15'', 17'' and $r_{\rm{s}}\,{=}\,$2'', 2.4''. The resulting combined posterior distribution is shown in Fig.\,\ref{fg:rht_kick}.



\bibliography{Pa30_GMOS}{}
\bibliographystyle{aasjournal}



\end{document}